# Wide-field spectral super-resolution mapping of optically active defects in hBN

*Jean Comtet[1,*], Evgenii Glushkov[1], Vytautas Navikas[1], Jiandong Feng[2,*], Vitaliy Babenko[3], Stephan Hofmann[3], Kenji Watanabe[4], Takashi Taniguchi[4], Aleksandra Radenovic[1,*]*

[1]Laboratory of Nanoscale Biology, Institute of Bioengineering, School of Engineering, École Polytechnique Fédérale de Lausanne (EPFL), 1015 Lausanne, Switzerland

[2] Zhejiang University, Tianmushan Road 148, Xixi Campus, Xi-6, 201, Hangzhou, 310027, China

[3]Department of Engineering, University of Cambridge, JJ Thomson Avenue, CB3 0FA Cambridge, United Kingdom

[4]National Institute for Materials Science, 1-1 Namiki, Tsukuba 306-0044, Japan

**Point defects can have significant impacts on the mechanical, electronic and optical properties of materials. The development of robust, multidimensional, high-throughput and large-scale characterization techniques of defects is thus crucial, from the establishment of integrated nanophotonic technologies to material growth optimization. Here, we demonstrate the potential of wide-field spectral single-molecule localization microscopy (spectral SMLM) for the determination of ensemble spectral properties, as well as characterization of spatial, spectral and temporal dynamics of single defects in CVD-grown and irradiated exfoliated hexagonal boron-nitride (hBN) materials. We characterize the heterogeneous spectral response of our samples, and identify at least two types of defects in CVD-grown materials, while irradiated exfoliated flakes show predominantly only one type of defect. We analyze the blinking kinetics and spectral emission for each type of defects, and discuss their implications with respect to the observed spectral heterogeneity of our samples. Our study shows the potential of wide-field spectral SMLM techniques in material science and paves the way towards quantitative multidimensional mapping of defect properties.**

**Introduction**

Defects in wide-band gap semiconductor materials lead to the creation of energy states well within the band gap, conferring these materials with new and exciting properties due to quantum confinement. A popular and well-studied example of point defects are nitrogen-vacancy (NV) centers in diamond, which can act as single-photon emitters[1], ultrasensitive magnetometers[2] and biological sensors[3].

In recent years, 2D semiconductors and insulators have emerged as new classes of materials able to host functional defects. Single-photon emission from such defects has been demonstrated in 2D Transition-Metal Dichalcogenides at cryogenic temperatures[4–7] and in hexagonal boron nitride (hBN) at room temperature[8]. Defects in 2D materials show great promise for many applications in integrated photonics[9,10]. The 2D nature of the material favors integration in photonic circuits, while allowing high light-extraction efficiency[10]. Defects in 2D materials also exhibit high sensitivity to their environment, leading to tunable properties[11] and allowing deterministic positioning of emitters through strain engineering[12–14].

However, defects can also be detrimental, leading to a decrease in the electrical and mechanical properties of materials. For example, the use of hBN as an encapsulation layer in nanoelectronics[15] can be dramatically affected by the presence of defects, causing increased scattering or facilitated dielectric breakdown[16], and precluding for now the use of CVD-grown hBN materials as efficient insulating layers in nanoelectronics[17].

Due to their atomic-scale nature, characterizing defects can be very difficult. Defects can be imaged with high spatial resolution in a transmission electron microscope (TEM)[18], but TEM imaging by itself tends to induce more defects in the sample, and is restricted to very small (~10 nm$^2$) areas. High-resolution Scanning Probe Microscopy techniques have allowed characterization of 2D materials at the single-defect level[19], and 2D material systems have also been investigated using near-field scanning optical microscopy (NSOM)[20]. Unfortunately, all these scanning techniques share similar limitation as TEM concerning time-consuming sample preparation and small imaging areas. Finally, confocal techniques[8] have been successful at characterizing the optical properties of single defects, but require very sparse samples due to diffraction-limited imaging. There is thus a clear demand in the development of new strategies allowing large-area characterization of defects in 2D materials.

With this goal in mind, we recently established single-molecule localization microscopy (SMLM) as a viable strategy for wide-field mapping of optically active defects in hBN[18]. Using the transient emission properties of individual emitters, we could separate emitters spatially down to 10 nm[18]. However, purely spatial SMLM techniques are still hampered by their lack of contrast, which can lead to substantial overcounting or undercounting of defects densities and precludes multidimensional measurement of defects properties, such as polarization, and spectra.

In this work, we focus on prototypical optically-active defects in hBN, and demonstrate that the implementation of spectral information concomitant to spatial SMLM (obtained by placing a prism in the detection path) is a valuable step forward in the large-area, nondestructive characterization of 2D materials and beyond. Our combined wide-field spectral and spatial super-resolved technique allows the determination of statistical ensemble spectral properties, as well as extraction of spatial, spectral and temporal dynamics of single defects. We demonstrate our approach on both monolayers of CVD-grown hBN materials, as well as irradiation induced surface defects in bulk exfoliated hBN materials. Similar operational principles could be applied to transition metal dichalcogenides at cryogenic temperatures[4–7] or other wide-band gap materials at room temperature such as aluminum nitride, silicon carbide and perovskites[21,22]. Our approach opens up broad perspectives in the use of SMLM techniques for wide field characterization of defects in a variety of materials, and paves the way for wide-field quantitative multidimensional imaging of optically-active defects[23].

**Experimental set-up**

Our set-up is based on a wide-field spectral super-resolution imaging scheme, using a prism in the detection path, previously developed in a biological context[24–26], and with recent extension in physical science fields such as chemistry[27]. As shown schematically in Figs. 1a-b, we excite the hBN samples using a 561 nm laser. The energy of the excitation laser (2.21 eV) is well below the band-gap of the hBN material (~ 6 eV), leading to the selective excitation of defects with energies well within the band-gap of the material. Excitation beam from the laser is focused onto the back focal plane of a high numerical aperture oil-immersion microscope objective, leading to the widefield illumination of the sample. hBN samples are deposited on silicon chips and immersed in water to prevent changes of refractive index in the optical path. Wide-field photoluminescence emission from the sample is collected by the same objective and separated from the excitation laser using dichroic and emission filters. Emission light is then split through two paths using a beam-splitter (Fig. 1b). Part of the emission going through Path 1 is directly projected onto one half of an EM-CCD chip, with a back-projected pixel size of 100 nm (Fig. 1d, Path 1). Emission from individual defects leads to the appearance of diffraction-limited spots on the camera chip (Fig. 1d, red boxes), which can then be localized with sub-nanometer accuracy, with a localization uncertainty $\sigma_{x,y} \sim \frac{\sigma_{PSF}}{\sqrt{N}}$, where $\sigma_{PSF} \approx$ 150 nm is the standard deviation of the gaussian fit of emitter's intensity (corresponding to a diffraction-limited spot fixed by the Point Spread Function with FWHM of $\approx$ 350 nm) and $N$ is the number of photons emitted by the defect during the acquisition of one frame. The second path consists of an equilateral calcium fluoride ($CaF_2$) prism. This dispersive element leads to an approximately linear shift $\Delta y_{SPEC}$ [px] in the photoluminescence signal of each emitters relative to their emission wavelength $\lambda$, such that $\Delta y_{SPEC} \sim a \times \lambda$, with $a \approx 0.25$ px/nm (see SI Fig. 1). As shown in Fig. 1e, the one-to-one correspondence between the simultaneous images of emitters in the spatial and spectral channels allows mapping of the spectrum of each individual emitter. The sample is further mounted on a piezoelectric scanner, and vertical drift is compensated using an IR-based feedback loop[28]. Residual lateral subpixel drift is compensated through postprocessing using cross-correlations between reconstructed super-resolved images (see Methods for further experimental and computational details).

**Results and Discussion**

Continuous green laser illumination of the flakes leads to photoswitching (blinking) of the emitters between bright and dark states[18]. As shown in Fig. 1d, due to this blinking behavior, only a small subset of emitters is active between each frame. This sparse activation allows for spatial localization of individual emitters between each camera frame, and subsequent reconstruction of a spatial and spectral map by summing up successive images, as presented in Fig. 1e.

As shown in Fig. 2, our wide-field spectral SMLM scheme allows us to map the spectral properties of emitters present in the hBN flakes. We first plot in Fig. 2a the distribution of center spectrum for a CVD-grown flake. As can be seen in this distribution, two families of emitters stand out clearly and are characterized by emission spectra centered approximately around $\lambda_1 \approx$ 585 nm ("green emitters") and $\lambda_2 \approx$ 640 nm ("red emitters"). We can further map the spatial position of each type of emitters in the reconstructed spatial maps of Fig. 2b, by summing up individual localization events (as depicted schematically in Fig. 1e). The brighter dots on the map thus represent the most active defects, which are emitting through most of the acquired frames. This spectral map allows for a direct estimation of the spatial localization of each defect type. Despite the presence of areas with varying densities of defects in each flakes, no clear spatial segregation is observed between red and green emitters, as they are homogeneously represented throughout the sample.

This type of multimodal spectral distribution is observed throughout our samples. We report in Fig. 2c the values of center wavelength for 5 distinct flakes, obtained using similar growth conditions and transfer procedures. Note that several modes can occasionally be observed for the red emitters (Flakes 2 and 4 in Fig. 2c, see Fig. S2 for details). Overall, emitters centered around 585 nm have relatively narrow spectral linewidth (FWMH ≈ 15 nm), consistent with recent report on similar CVD-grown materials[29], while the second group of emitters (with wavelength between approximately 610 nm and 650 nm) have larger spectral linewidth and show a relatively large sample-to-sample variation, which can be attributed to a variation in the local mechanical[30,31] and electrostatic[29] environment associated with each flake (e.g. due to residual strains developed during the transfer process). Finally, we show in Figs. 2d-e the evolution of the number of localizations per frames and the spectral distribution over more than 30 minutes. As is depicted in Figs. 2d-e, the number of localizations per frames decreases exponentially, with a characteristic time $\tau \approx 10$ s due to bleaching of the emitters, while spectral emission remains stable, with progressive reduction of the FWMH of red emitters probably due to bleaching. Importantly, localizations and wavelength matching can be performed as fast as 20 per frames, showing the potential of the technique for large-area and high-throughput mapping in dense samples. Bleaching in turn can happen due to irreversible photo-oxidation of defects, exposed to the external environment due to the 2D nature of the flakes.

We now proceed in Fig. 3, to the characterization of the spatial, spectral and temporal dynamics of individual emitters in hBN, using our wide-field SMLM spectral and spatial super-resolution scheme. We show that at least two types of defects are responsible for the observed emission lines in CVD-grown flakes (Figs. 2a-c). In Fig. 3a we first report super-resolved images of individual emitters, rendered as spatial histograms of localizations, using a pixel size of 5 nm. As shown in Fig. 3b, we obtain approximately gaussian distributions, corresponding to spatial uncertainty $\sigma_X$ between 15 and 20 nm. This value is slightly larger than the estimated spatial uncertainty of each localization, based on photon counts (of the order of 7 nm), which is probably due to residual drift. Remarkably, over the entire number N ≈ 1000 of localizations, this leads to a corresponding final uncertainty for the individual defect center position as low as $\frac{\sigma_X}{\sqrt{N}} < 1$ nm for the most active defects. The central emission wavelength of individual defects (Fig. 3c) and full emission spectrum (Fig. 3d) can then be measured using the procedure described in Fig. 1d. As shown in Fig. 3c, for the two representative types of emitters, relatively constant spectral emission is observed over 600 successive localization frames, with distribution width $\Delta\lambda \approx 5$ nm.

Representative spectra for the first defect type ("type A") are shown in green in Fig. 3d. These defects have a mean emission wavelength centered along the first peak $\lambda_1 \sim 585$ nm in the spectral histogram of Figs. 2a and 2c. Emission spectrum is asymmetric and a phonon sideband can be occasionaly resolved on some spectra (Fig. 3d, green arrow). The second type of defect ("type B") is characterized by a mean emission wavelength centered along the second peak $\lambda_2 \sim 610 - 650$ nm of the spectral histogram. The spectrum shows a clear Zero Phonon Line and Phonon Sideband, leading to an energy detuning of ~ 140 meV, consistent with previously published work[32–34]. Remarkably, this phonon sideband is also visible on the ensemble histograms, in the form of a third local maximum in the spectral distribution (Fig. 2a, red arrow). A representative time trace for the emission of each individual emitter is shown in Fig. 3e. All defects systematically exhibit a blinking behavior, characterized by emission intermittency, and successive ON and OFF events.

To verify the generality of our approach, we characterized defects in a second class of hBN materials. We started from high-quality bulk hBN crystals[35], which were exfoliated and deposited on Si/SiO$_2$ substrate (see Fig. S3). Few emitters, corresponding to both type A and type B, were observed on the as-exfoliated flakes, traducing the high quality of the bulk hBN material (see Fig. S4). In order to deterministically induce defects in the structure, we exposed the exfoliated crystals to 30 s of oxygen plasma treatment[36] (see Materials and Methods). As shown in Fig. 4a, this leads to creation of emitters at the surface of the flakes, with similar blinking behavior as in CVD-grown materials, allowing straightforward localization using SMLM-based spectral super-resolution mapping (representative time trace shown in Fig. S5). Interestingly, as illustrated in Figs. 4b-c, the spectral distribution in most of the investigated samples is characterized by a single emission wavelength, showing preferential creation of type A emitters by plasma treatment. Surprisingly, some of the investigated flakes also showed a significant population of type B emitters (Fig. 4c, flakes number 2 and 6 and Fig. S6) although most of these emitters are unstable and bleach irreversibly after a few tens of seconds. This differential bleaching could be attributed to distinct chemical reactivity of the two populations of defects.

Focusing on CVD-grown hBN, we now turn in Fig. 5 to the photophysical properties of emitters and specifically their intensity and blinking kinetics. We show in Fig. 5a a 2D histogram of photon counts as a function of the emission wavelength for the flake investigated in Fig. 2. Two clusters (green and red dashed circles) corresponding to type A and type B defects identified in Figs. 2a and 3 stand out clearly, with high brightness of the order of $\sim 5.10^4$ photons/second. As evidenced on the map, no clear difference of brightness can be made between each type of defects. Phonon Sideband for type A and type B emitters can be identified in this histogram (black circled clusters). To obtain more insight into the properties of the defects in terms of photon emission, we show in Fig. 5b the full distribution of emission intensity by grouping the defects according to their wavelength. Interestingly, we observe a long tail in the intensity distribution, in strong contrast to the poissonian distribution classically expected for non-blinking emitters[37].

Another interesting feature of the emitters investigated here is their blinking kinetics, also observed in several other studies[18,34,37], but contrasting with other reports of long-term emission stability for the emitters in hBN[10,36]. To gather more insight into this blinking behavior, we plot in Figs. 5c-d the distributions of ON and OFF time for defect types A and B (see Materials and Methods). We observe clear power-law distributions, with $P_{on/off}(t) \sim t^{\alpha_{on/off}}$, and $\alpha_{on} \approx \alpha_{off} \approx 1.9 \pm 0.2$ (red and green lines, Figs. 5c-d). Remarkably, these power-law distributions and the associated exponents are consistent with previously observed blinking kinetics on quantum dots[38,39], where blinking is attributed to ionization of the quantum dots and escape of photoexcited carriers towards surrounding charge traps in the material. Importantly, the power-law scaling characterizes the absence of intrinsic time-scales in the blinking behavior and can be interpreted as being due to heterogeneity of charge traps at the material's surface. Similar mechanisms due to photoinduced ionization and change of the charge state of the defects were proposed to explain blinking in NV centers in diamond[40]. The blinking behavior observed in our samples might thus take its origin from charge separation or ionization of the defects, as the monolayer nature of the CVD-grown hBN flakes, and the creation of surface traps in plasma-treated hBN crystals, might allow in both cases for a facilitated escape of photoexcited charge carriers towards surrounding charge traps. This observation suggests a rationale behind the increased photostability observed in annealed samples, as a way to desorb impurities acting as charge traps. Remarkably, the observed similarity in blinking kinetics suggests similar charge affinities for each types of defects (red and green dots, Figs. 5c-d). Finally, the long tail in the distribution of emission intensity, shown in Fig. 5b, could be related to the observed power-law distribution of blinking times.

An important question remains to understand the reasons for the spectral heterogeneity evidenced in our study and the presence of different emission lines and defects states in CVD grown hBN and irradiated bulk hBN crystals. In the quest towards the assignment of a precise chemical structure to optically active defects in hBN materials, recent works have shown a widely heterogeneous spectral response of defects in hBN[32,41], which was attributed to different chemical structures[42,43], but also differences in the charge states[34,44], local dielectric[33], electrostatic[29] and strain environment[30], as well as temperature[41]. Multimodal emission lines with modes around 585 and 630 nm were also observed in several studies[34,43]. While the blinking behavior observed for each types of defects (Fig. 3e) is probably due to reversible ionization or charge trapping, it is also concurrent with clear spectral stability for the emission in the ON state (Fig. 3c). Furthermore, the observation of similar blinking characteristics for types A and B defects (Figs. 5c, d) suggests similar charge affinities for these two populations. The differences in the spectral shape and zero phonon line distribution of type A and B defects (Figs. 2c and 3d) could thus point to distinct chemical structures for these two populations, variations in their local dielectric environment or distinct charge states. Furthermore, the large variation in spectral emission amongst type B defects within different CVD-grown flakes might be attributed to residual strain developed during the transfer process. Despite their relatively large uncertainty, we can expect theoretical predictions from Density Functional Theory to further guide the identification of the exact chemical structure of each type of defects[45], based on their emission wavelength and spectra. Noteworthy, Adbi and coworkers[46] reported Zero Phonon Line for $N_BV_N$ and $V_B^-$ defects of 2.05 eV and 1.92 eV, respectively, corresponding to 605 nm and 647 nm emission wavelength, which is in relatively fair agreement with our experimental results. However, the fact that we find predominantly one defect type in plasma treated exfoliated flakes might also suggest that type A correspond to intrinsic defects, while type B might occur due to substitutional doping with various impurities, such as $C_BV_N$[47], or to various oxygen or hydrogen complexes or interstitial defects[48]. Critically, the ability to assign defined chemical structures to optically-active defects in hBN will only be possible through studies of the statistical properties of the spectral response of defects submitted to various growth conditions and post-growth treatments.

**Conclusion**

We have shown the potential of spectral SMLM techniques for wide-field and high-throughput spectral mapping and characterization of defects in hBN materials. Our methodology allows the determination of statistical ensemble spectral properties, as well as extraction of spatial, spectral and temporal dynamics of single defects. We identify at least two types of defects in CVD grown materials, while irradiated exfoliated flakes show predominantly one type of defect. Further analysis of the blinking kinetics of optical emitters suggests that blinking is due to transient trapping of photoexcited charge carriers at the material's surface, similar to what is observed in quantum dots. This behavior provides strategies for blinking reduction and increase of photostability through encapsulation of defects and reduction of charge traps via thermal annealing. Our study demonstrates the potential of spectral SMLM as a wide field and high-throughput characterization technique in material science and paves the way towards multidimensional mapping of defects' properties.

**Materials and Methods**

*Optical set-up*

The sample is excited using a 561 nm laser (Monolitic Laser Combiner 400B, Agilent Technologies). Excitation power during imaging, measured at the back focal plane of the microscope objective varies from 20 to 65 mW, corresponding to power densities ranging from 250-800 kW/m$^2$. We used a high numerical aperture oil-immersion microscope objective (Olympus TIRFM 100X, NA = 1.45). Wide-field photoluminescence emission from the sample is collected by the same objective and separated from the excitation laser using dichroic and emission filters (ZT488/561rpc-UF1 and ZET488/561m, Chroma). Paths 1 and 2 consist of two telescopes[26], sharing the same lens L1, with a respective magnification factor 1.6 (Path 1, telescope L1-L2) and 1.4 (Path 2, telescope L1-L3). Lenses are achromatic doublet lenses (Qioptic, L1: f/100, L2: f/160, L3: f/140). The prism (PS863, Thorlabs) is placed at the Fourier plane between L1 and L3, at the angle of minimum deviation. The sample is mounted on a piezoelectric scanner (Nano-Drive, MadCityLabs) to compensate for vertical drift using an IR-based feedback loop[28]. EM-CCD camera (Andor iXon Life 897) is used with an EM gain of 150.

*Fitting and spectral assignment and SMLM images*

*Localization:* Emitters in the spatial and spectral channels are localized using the imageJ plugin, *Thunderstorm*[49]. Briefly, a wavelet filter is applied to each frame. Peaks are then fitted by 2D integrated Gaussians. In the spatial channel, only emitters with intensity at least twice of the background are considered.

*Spectral calibration:* Calibration between spatial and spectral channels is conducted using red fluorescent beads (20 nm FluoSpheres carboxylated-modified, 580/605). First, we calibrate the field-of-view transformation between spatial and spectral channels. For an emitter with a given wavelength, the relation between its position in the spatial and spectral channel is well approximated by a relation of the form $(x_{SPEC}, y_{SPEC}) = A \cdot (x_{LOC}, y_{LOC}) + B$, where $(x_{SPEC}, y_{SPEC})$ and $(x_{LOC}, y_{LOC})$ corresponds to the position vector of the emitter in the spectral and spatial channels, respectively, A is a 2x2 matrix and B is a vector. Coefficients for A and B are calibrated by raster scanning a fiducial marker of controlled emission wavelength and mapping its position in both channels. In a second step, bandpass filters with a central wavelength of 600 nm, 630 nm and 650 nm, respectively, and FWHM of 10 nm (Thorlabs, FB600-10) are used to calibrate the relation between the emission spectrum and the vertical shift in the spectral channel. The relation between the vertical shift $\Delta y_{SPEC}$ [px] of the emission spectra and the emission wavelength $\lambda$, is well approximated by a linear relation such that $\Delta y_{SPEC} \sim a \times \lambda$, with $a \approx 0.25$ px/nm (See Fig. S1).

*Spectral assignment:* In order to assign a spectra to an emitter, localized in the spatial channel at $(x_{LOC}, y_{LOC})$, we compute its projected position $(x_{SPEC}(\lambda_0), y_{SPEC}(\lambda_0))$ in the spectral channel for a fixed emission wavelength $\lambda_0 = 650$ nm. A pair-search algorithm finds the closest localizations $(x'_{SPEC}, y'_{SPEC})$ in a vertically-elongated rectangular zone around $(x_{SPEC}(\lambda_0), y_{SPEC}(\lambda_0))$. The corresponding central emission wavelength $\lambda$ is then estimated as $\lambda = \lambda_0 + 1/a \cdot (y_{SPEC}(\lambda_0) - y'_{SPEC})$.

*Spectra generation:* Measured spectra in Fig. 3d are obtained by averaging the spectrum of single emitters over all frames for which spectral assignment is successful.

*Image generation:* SMLM images (Fig. 2b and 4a) are generated as probability maps by plotting 2D Gaussian centered on each position with a standard deviation equal to 20 nm.

*Drift correction:* Lateral drift is corrected using cross-correlation between reconstructed super-resolved images on CVD-grown materials. No lateral drift correction is applied on exfoliated hBN flakes, as the lower density of defects leads to weak cross-correlations between reconstructed images.

*Blinking kinetics:* We obtain the blinking kinetics of type A and type B defects (Fig. 5c-d) by following the time-trace of 9 type A defects and 14 type B defects (histogram of mean emission wavelength shown in Fig. S7). OFF state is clearly defined by the absence of localization (Fig. 3e).

***Sample preparation***

*CVD-grown materials:* CVD-grown materials are produced under similar conditions as described elsewhere[50] (see Fig. S8 for SEM images of flakes on Fe foil). Briefly, as-received Fe foil (100 µm thick, Goodfellow, 99.8% purity) is loaded in a customized CVD reactor (base pressure $1\times10^{-6}$ mbar) and heated to ~940 °C in Ar (4 mbar), followed by annealing in $NH_3$ (4 mbar). For the growth, $1\times10^{-2}$ mbar $NH_3$ is used as a carrier gas and $6\times10^{-4}$ mbar borazine $(HBNH)_3$ is introduced into the chamber for 30 minutes. The growth is quenched by turning off the heater allowing a cooling rate of about 200 °C/min.

The h-BN domains are transferred onto $SiN_x$ chips using the electrochemical bubbling method[51] with PMMA as support layer. After transfer, PMMA is removed by successive 1-hour rinses in hot acetone (3 rinses), hot IPA (1 rinse) and hot DI water (1 rinse). Remaining PMMA contamination is further removed through overnight annealing at 400°C in an Argon atmosphere.

*Exfoliated flakes:* hBN multi-layer flakes are exfoliated from high quality bulk crystals[35] and deposited onto $SiO_2$ chips (Fig. S3). Type A and type B defects are detected at low concentrations in just-exfoliated flakes (Fig. S4). In order to deterministically create defects, hBN crystals are further exposed to a 30 s oxygen plasma at 100 mW under 30 sccm $O_2$ flow. No annealing was performed.

*Sample imaging:* Chips with deposited hBN flakes and CVD-grown materials are placed upside down on round coverslips (#1.5 Micro Coverglass, Electron Microscopy Sciences, 25 mm in diameter), previously cleaned in oxygen plasma for 5 minutes. Imaging is further performed in water at room temperature, in order to improve the optical contrast and prevent a discontinuous change in the refractive index. Performing experiments in air by transferring CVD-grown hBN flakes directly on a glass coverslip, we observed a similar spectral signature as that shown in Fig. 2, showing that water has little effect on the observed spectral response of this material.

**Supporting Information.**

Supplementary Figures S1-S8 (PDF).

**Corresponding Author**

*jean.comtet@gmail.com, jiandong.feng@zju.edu.cn, aleksandra.radenovic@epfl.ch

**Author Contribution**

AR and JF conceived the project and designed experiments. JC developed the experimental set-up for spectral imaging and performed the experiments, with help from EG, NV. VB and SH produced the CVD-grown hBN samples. KW and TT produced the bulk hBN crystals. JC wrote the paper, with input from all authors. AR supervised the project.

**Funding Sources**

This work was financially supported by the Swiss National Science Foundation (SNSF) Consolidator grant (BIONIC BSCGI0_157802) and CCMX project ("Large Area Growth of 2D Materials for device integration"). K.W. and T.T. acknowledge support from the Elemental Strategy Initiative conducted by the MEXT, Japan and the CREST (JPMJCR15F3), JST.
V.B. and S.H. acknowledge funding from the European Union's Horizon 2020 research and innovation program under grant agreement No number 785219.

**ACKNOWLEDGMENT**

We would like to thank Jean-Baptiste Sibarita and Corey Butler for the initial help and experiments on spectral SMLM. We also acknowledge valuable discussions with Adrien Descloux and Kristin Grussmayer. We thank Ahmet Avsar for his help with the transfer of exfoliated hBN, Michael Graf for fabricating marked silicon substrates, and Ivor Lončarić for the discussion on the interpretation of our results.17

FIGURES

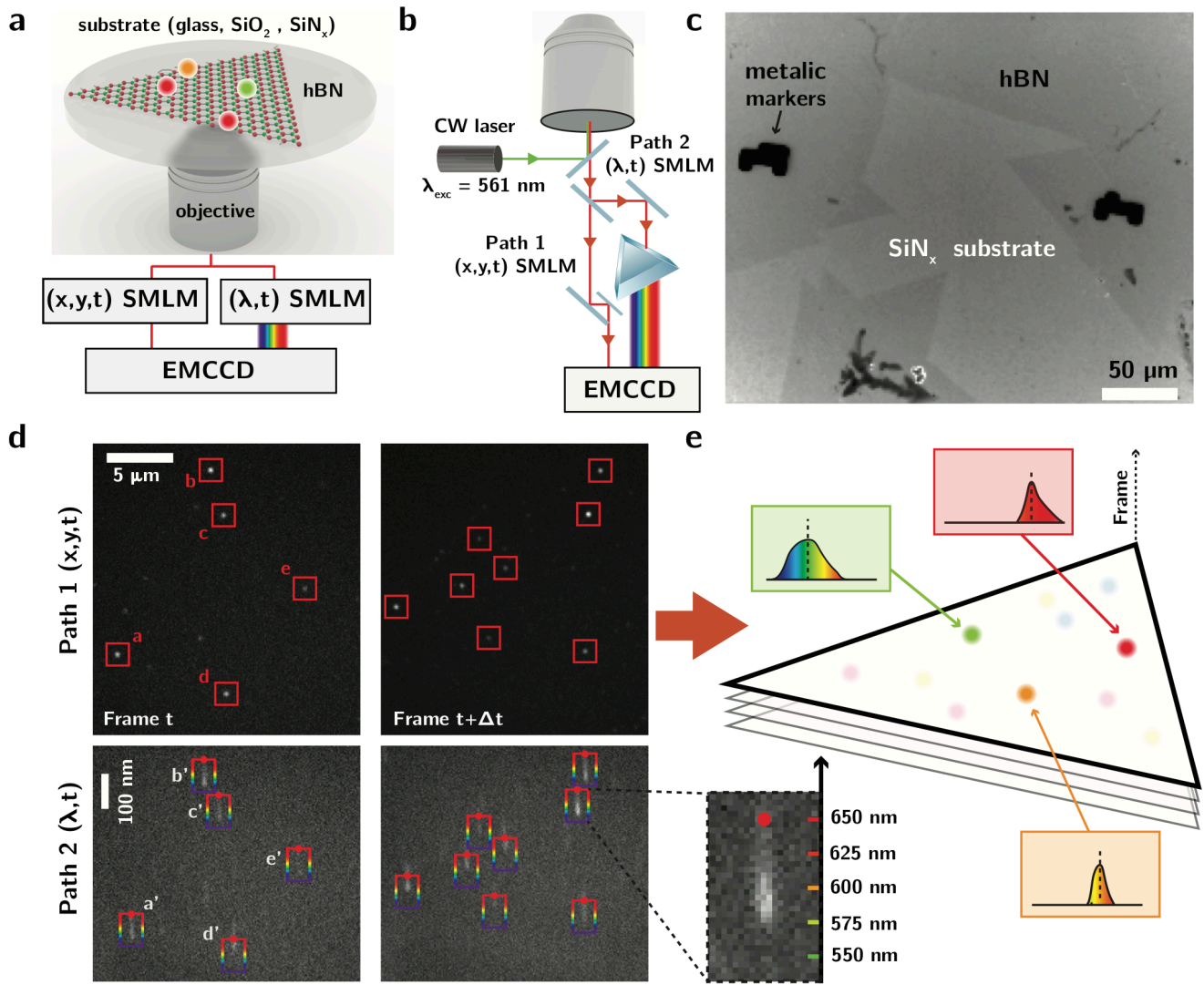

**Figure 1: Ultrahigh-throughput prism-based wide field spectral characterization of optical emitters in hBN. (a)** Principle of the experimental set-up, allowing spatial and spectral Single Molecule Localization Microscopy (SMLM) of emitters in hexagonal boron nitride (hBN) materials deposited on various substrates (glass, $SiO_2$ and $SiN_x$ chips). **(b)** Schematic of the experimental set-up. Upon laser excitation, fluorescence signal emanating from single defects is collected by a high NA objective and split towards two distinct paths for spatial (path 1) and spectral (path 2) characterization. Spatial Path 1 leads to diffraction-limited spots for individual emitters, which can be localized with sub-pixel accuracy. Spectral path 2 is

composed of a dispersive prism element, shifting the fluorescence of individual emitters according to their emission wavelength. Images from both paths are then projected on the same chip of an EMCCD camera (see d). **(c)** Wide field image of CVD grown hBN flakes transferred on Si/SiN$_x$ chips. **(d)** Simultaneously recorded wide field image (path 1; x,y,t) and spectral image (path 2; λ,t) of emitters in exfoliated hBN flakes between successive frames at t and t+Δt. Red boxes in Path 1 indicate spatial position of individual emitters, corresponding to diffraction-limited spots, localized with subpixel accuracy. Multicolor boxes in Path 2 show the corresponding images of individual emitters after vertical dispersion by the prism element. Red dots in spectral channel indicate the mapped spectral position of 650 nm for each emitter in the spatial channel. The one to one correspondence between spatial and spectral path allows to obtain the full spectra of each individual emitters, through measurement of the vertical shift in the spectral channel (see zoom, Fig. d). Images are averaged over 5 frames for clarity (with 20 ms exposure). **(e)** Principle of the reconstructed spectral super-resolved image. Spatial positions of emitters, along with their attributed spectra are summed up over successive frames, allowing the reconstruction of a super-resolved map.

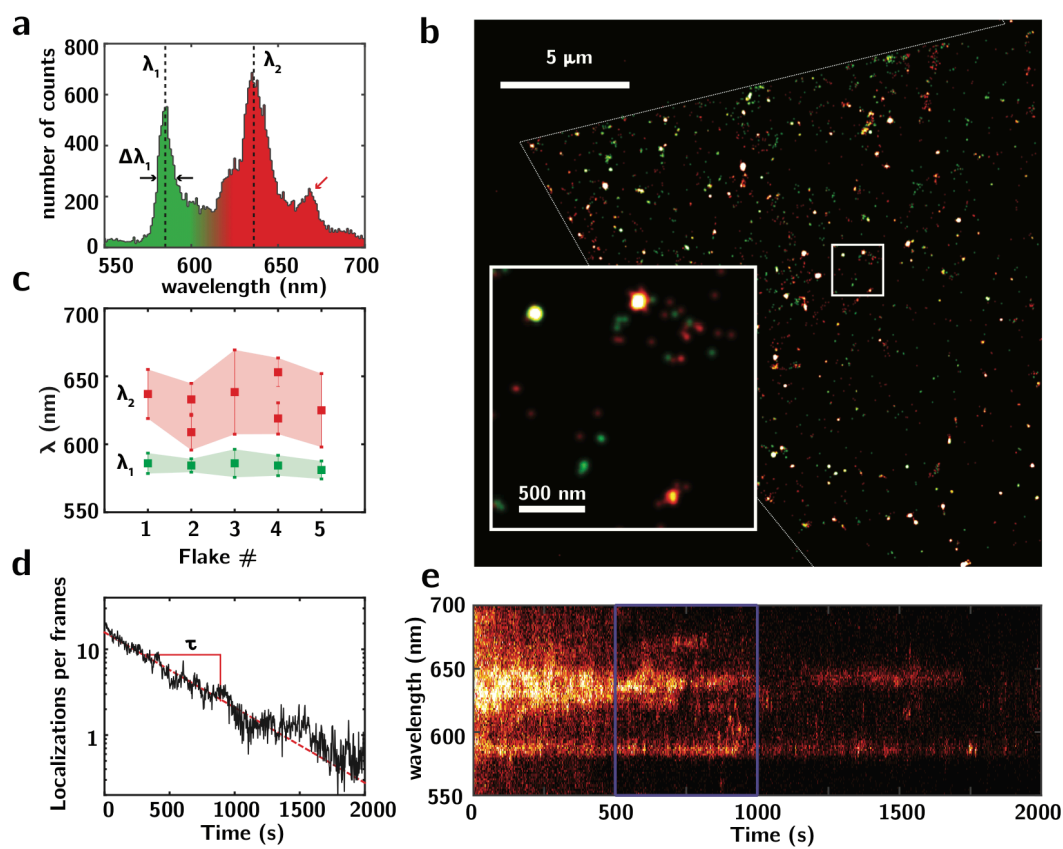

**Figure 2: Wide field super-resolved spectral and spatial map of single emitters in CVD-grown hBN flakes. (a)** The spectral distribution of center emission wavelengths for a CVD-grown hBN flake. A dual distribution is observed, corresponding to green and red emitters of respective center emission wavelength $\lambda_1$ and $\lambda_2$. Smaller emission peak indicated by red arrow corresponds to a phonon side-band (see text for details). Bin size is 1 nm. **(b)** Corresponding spatial map with spectral contrast, showing the position of the two types of emitters throughout the flakes. Localization events are rendered as gaussian spots (see Methods). Spatial maps are rendered from 10,000 successive frames. **(c)** Sample to sample variation of center wavelengths. Error bars are FHWM. **(d)** Temporal evolution of the number of localizations per frames (acquisition with 50 ms sampling and 20 ms exposure time). **(e)** Temporal evolution of the spectral distribution. Blue zone indicates the region used for reconstructing the spatial images in (b) and the distribution in (a).

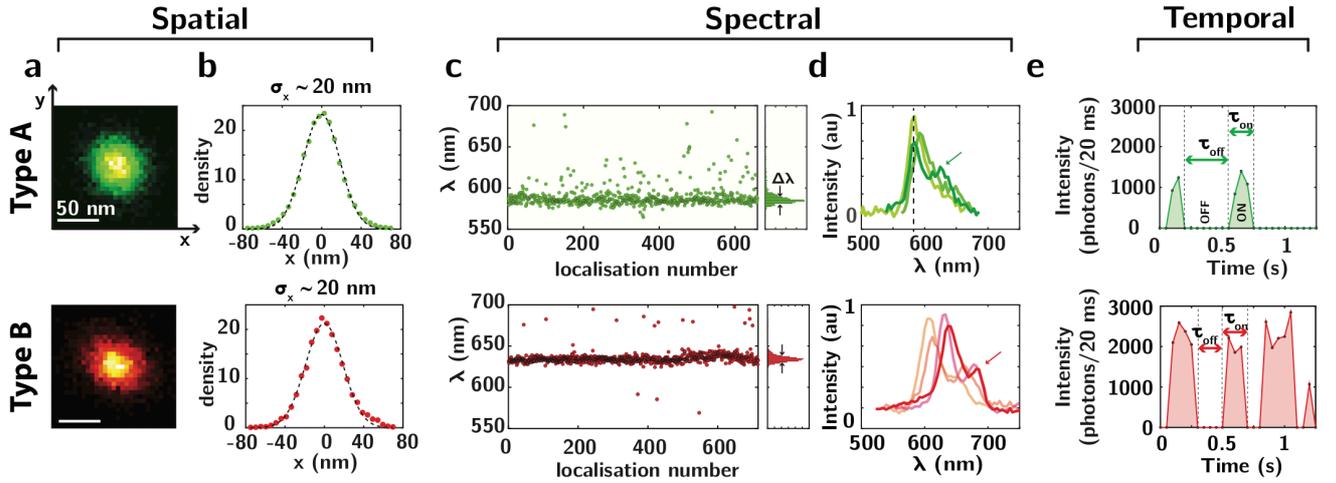

**Figure 3: Spatial, Spectral and Temporal dynamics for the two types of defects.** (a) Super-resolved image of individual emitters, rendered as spatial (2D) histogram of localization coordinates, with 5 nm pixel size. (b) Projected distribution along the x axis along with gaussian fit (dotted line), allowing the determination of localization uncertainty $\sigma_x$. (c) Variation of the mean emission wavelength over successive localizations, and histogram of emission wavelength. (d) Representative spectra of emitters. The vertical dashed line represents mean emission wavelength. Green and red arrows show respectively phonon sideband in type A and type B emitters. (e) Time trace of emission intensity, showing intermittency in the emission, leading to successive ON and OFF states. All measurements presented here are acquired with 50 ms sampling and 20 ms exposure time.

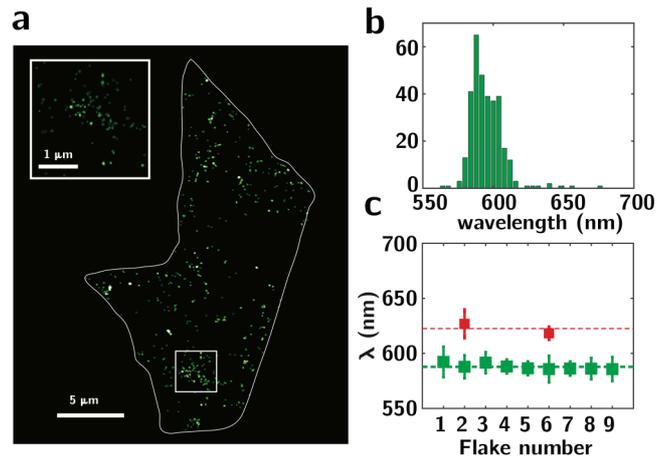

**Figure 4: Spatial and spectral characterization of plasma treated exfoliated hBN flakes.** **(a)** Reconstructed map of individual emitters in exfoliated hBN flake, submitted to oxygen plasma for 30 s (see Methods). **(b)** Distribution of emission wavelength. **(c)** Sample to sample variation of the emission wavelength. Bin size is 4 nm.

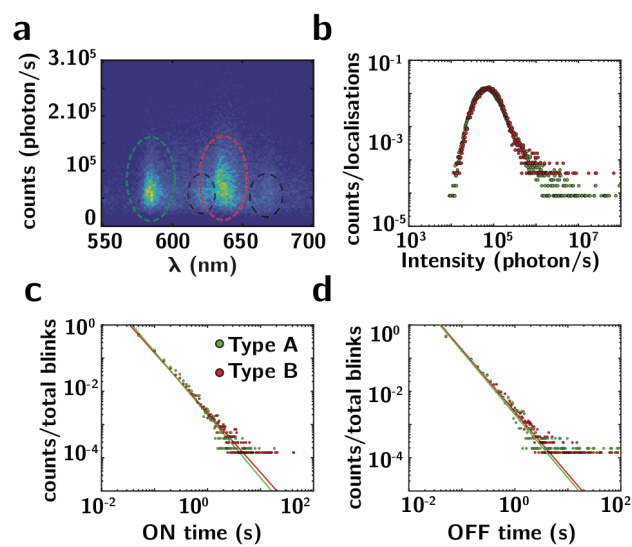

**Figure 5: Photo physical properties of emitters in CVD hBN. (a)** 2D histogram of photon counts as a function of emission wavelength. The two main clustered distributions are circled in green and red. Phonon Sideband of type A and type B emitters are circled in black. **(b)** Histogram of emission intensity. Red and green dots correspond to each circled distribution in (a). Normalized histograms of on **(c)** and off **(d)** times for type A and type B emitters (see Supplementary and Methods). Straight lines indicate power-law scaling of each distribution.